\documentclass[twoside]{article}

\usepackage{lipsum} % Package to generate dummy text throughout this template

\usepackage[sc]{mathpazo} % Use the Palatino font
\usepackage[T1]{fontenc} % Use 8-bit encoding that has 256 glyphs
\usepackage{microtype} % Slightly tweak font spacing for aesthetics

\usepackage[hmarginratio=1:1,top=32mm,columnsep=20pt]{geometry} % Document margins
\usepackage{multicol} % Used for the two-column layout of the document
\usepackage[hang, small,labelfont=bf,up,textfont=it,up]{caption} % Custom captions under/above floats in tables or figures
\usepackage{booktabs} % Horizontal rules in tables
\usepackage{float} % Required for tables and figures in the multi-column environment - they need to be placed in specific locations with the [H] (e.g. \begin{table}[H])
\usepackage{hyperref} % For hyperlinks in the PDF

\usepackage{lettrine} % The lettrine is the first enlarged letter at the beginning of the text
\usepackage{paralist} % Used for the compactitem environment which makes bullet points with less space between them

\usepackage{abstract} % Allows abstract customization
\usepackage{titlesec} % Allows customization of titles
\renewcommand\thesection{\Roman{section}} % Roman numerals for the sections
\renewcommand\thesubsection{\Roman{subsection}} % Roman numerals for subsections
\titleformat{\section}[block]{\large\scshape\centering}{\thesection.}{1em}{} % Change the look of the section titles
\titleformat{\subsection}[block]{\large}{\thesubsection.}{1em}{} % Change the look of the section titles

\usepackage{fancyhdr} % Headers and footers
\title{\vspace{-15mm}\fontsize{24pt}{10pt}\selectfont\textbf{More network science for teenagers}} % Article title

\author{
\large
\textsc{Angel S\'anchez\thanks{Grupo Interdisciplinar de Sistemas Complicados (GISC). Website: http://www.anxosanchez.eu}\  \ \& Cristina Br\"andle}\\[2mm] % Your name
\normalsize Departamento de Matem\'aticas, Universidad Carlos III de Madrid \\ % Your institution
\normalsize 28911 Legan\'es, Madrid, Spain \\ 
\normalsize \href{mailto:anxo@math.uc3m.es}{anxo@math.uc3m.es}, % Your email address
\normalsize \href{mailto:cbrandle@math.uc3m.es}{cbrandle@math.uc3m.es} 
\vspace{-5mm}
}
\date{}

\begin{document}

\maketitle % Insert title

\thispagestyle{fancy} % All pages have headers and footers

%----------------------------------------------------------------------------------------
%	ABSTRACT
%----------------------------------------------------------------------------------------

\begin{abstract}

\noindent Recently, Harrington et al.\  (2013) presented an outreach effort to introduce school students to network science and explain why researchers who study networks should be involved in such outreach activities. Based on the modules they designed and their comments on the success and failures of the activity, we have carried out a sequel with students from a high school in Madrid, Spain. We report on how we developed it and the changes we made to the original material. 

\end{abstract}

%----------------------------------------------------------------------------------------
%	ARTICLE CONTENTS
%----------------------------------------------------------------------------------------

\begin{multicols}{2} % Two-column layout throughout the main article text

\section{Introduction}

\lettrine[nindent=0em,lines=3]{R} ecently, a group of researchers at the University of Oxford published a commentary \cite{mason} in which they describe the activities they have been carrying out to introduce teenagers to Network Science. In their paper, they state: 
\begin{quotation}
We are continuing to conduct outreach activities across England, but the only way
to make a really big impact is if these activities spread far and wide. We hope
that we have whet appetites for conducting outreach activities in schools, and we
encourage people to use, borrow, adapt, and improve any material in this article.
\end{quotation}

Inspired by their efforts, we have adapted the material provided by the Oxford group for using it with a high school in Madrid, Spain. Unfortunately, we have not had the time to repeat the outreach activity in different places, but we feel that the experience we had is enough for us to share the material we prepared and our experiences as to how the activity developed. 

%------------------------------------------------

\section{Methods}

We worked along with the Mathematics Department of the Instituto de Ense\~nanza Secundaria (High School) ``Blas de Otero'', in a suburb of Madrid called Aluche. We first presented the subject of complex networks to the teachers and discussed with them whether they found it interesting and appealing for the students, and whether it would be a good activity for the students to appreciate both the knowledge on networks and the importance of mathematics to understand them. We also checked at what level the exposition could proceed and decide on the type of activity we would carry out. We eventually decided on two types of activities:
\begin{compactitem}
\item an activity at our University for students aged 15 and 16 (4$^{\circ}$ ESO, equivalent to Year 11 in England), and
\item an activity at the School, for students aged 16 through 18 (1$^{\circ}$--2$^{\circ}$ Bachillerato, equivalent to Years 12 and 13 in England).
\end{compactitem}
We now describe the two types of activities.

%------------------------------------------------

\subsection{Activity at the University: Talk + workshop}

In this activity, groups of students were taken to the University by their teachers. We conducted two sessions on separate days, one with 50 students and the second one with 38 students (including some who receive special attention under a program called "diversificaci\'on" (diversification) to help them keep up with the level they should have). We agreed with the teachers that they would be probably more excited about the activity if it took place at the University and that it would be nice for them to see what the University looked like. The groups came to the University by public transportation. 

After guiding them to one of the lecture halls at the University (suited to the group size, not too big) we gave them a very brief introduction about the University (5 min) and then proceeded immediately with a presentation on Complex Networks and how they are present in many aspects of our lives and of our world. The presentation is available in the Web \cite{web} (note that it includes both the presentation and the subsequent workshop material, see below; original Keynote presentation is available upon request). The presentation lasted about 45 minutes and we encouraged the students to ask any questions they might have, insisting that we would rather cover less material than risk not being understood. During the presentation, we tried to give examples the students could be familiar with, referring many of the concepts to the context of the internet, for instance. Then we made a break (5 min) during which we asked the teachers to divide the group in two, which they did according to the original High School classes. One of the groups remained in the same lecture hall and the other one proceeded to an adjacent one. 

Then, we went through the second part of the talk, which was intended to be a workshop, in which the results were worked out among presenter and students. To this end, students were asked to bring their pens and we provided them with paper. The workshop was based on the ideas in \cite{mason} but instead of focusing on one or two of the topics as they did, we tried to give the students a tour through all of them. The workshop covered the following topics: 

\begin{compactitem}
\item Network structure: building a network from the raw data, graph isomorphism, connected graphs, cycles, diameters and clustering
\item Node importance: degree, betweenness and closeness centralities
\item Node importance: PageRank, iterative calculation, random walk calculation
\item Small world and six degrees of separation
\item The friendship paradox, or why my friends have more friends than I, and the effect of weighted averages
\item The configuration model
\item Communities and the Girvan-Newman algorithm
\item Structural balance and how it relates to social or military conflicts such as WWI
\item Epidemic spreading on different types of networks
\item Planar graphs and Euler's theorem
\item Ecological networks and the role of the different species
\end{compactitem}

This second part was carried out in a more interactive manner, asking questions to the students and waiting for them to come up with answers that were presented to and discussed with the groups. We paused as the workshop progressed to make sure they did try to look into the answers. In some cases, such as the application of the Girvan-Newman algorithm, we gave them the lists of the shortest paths to save time. 

\subsection{Activity at the High School}

The activity at the High School was programmed to be the lecture from an outside speaker in the Day of Science, and it involved approximately 100 students from 1$^{\circ}$ Bachillerato and those in 2$^{\circ}$ that take Mathematics. The activity took place at the main lecture hall of the High School, and it was almost full, this being another reason for not having all Bachillerato students in the activity. 

In this case, the presentation proceeded without interruption. The introduction to the University was somewhat longer (10 min) and then the presenter went through the two parts of the presentation without break. During the workshop part there were a few questions posed to the audience but the level of interaction was of course much smaller than in the activity at the University. 

%------------------------------------------------

\section{Results}

Globally, the two activities turned out quite well. For the students of 4$^{\circ}$ ESO, some teachers conducted anonymous polls among their classes and the responses were all A or B. Informal conversations with some of the students after both activities supported also this impression. It appears that the activities succeeded in transmitting many of the basic concept of network science and how mathematics helps us understand their relevance to our lives. From a general viewpoint, we are quite satisfied with how things went. 

Going now into the details of the activities, there are a number of points worth commenting:
\begin{compactitem}
\item The length of the first part is too much for the activity with the youngest students at the University. There are many examples and some of them are simply too far from what they know or are familiar with. After the first session we already noticed this problem so for the next day we prepared a somewhat shorter version of the talk. Both versions are available at the website for comparison. The second day we noticed an improvement of this issue: the shortened lecture did work better with the students. For the activity at the High School, the format being closer to a usual conference, the long introduction was not a problem. 
\item  The workshop is also too long. In the University sessions we were not able to get to the Epidemic spreading part in 1 hour, and that going a little bit too fast through the material, which made following it a little bit demanding for some of the groups. A decision needs to be made beforehand as to whether the session should be longer (probably not) or being more selective with the topics. 
\item The order of the topics is also not optimal. We found that beginning with simple questions about graph structure worked well, but then the students found the discussion of centrality concepts a little arid. We had though of using a graph of relationships in XVth century Florence to find that the Medici were the most central family, but discarded it because it was too complicated a graph and it would have taken too long to compute centralities. However, we now feel that it would have been better to make a less abstract discussion. There is room for improvement here. 
\item We never discussed the configuration model. We did not have nice examples for it, it was always something a bit complicated and given that we were running short of time, we skipped it (both presenters took the decision independently). A good pre-worked example of that would be needed to enter that topic. 
\item Most of the topics we covered were well received and attracted the students' attention. 
\item While the size of the two groups for the workshop session was manageable, it would have worked better with a third person so we could have formed three groups, or else with two people per group so discussions could be followed more closely by the presenters. 
\item In the High School, lecture format session, given that the amount of time was even shorter, the presenter went very quickly through the centrality part and skipped the structural balance part altogether. However, the quick pace helped the students (who were old enough) to stay focused and avoid getting bored. 
\end{compactitem}

%\begin{table}[H]
%\caption{Example table}
%\centering
%\begin{tabular}{llr}
%\toprule
%\multicolumn{2}{c}{Name} \\
%\cmidrule(r){1-2}
%First name & Last Name & Grade \\
%\midrule
%John & Doe & $7.5$ \\
%Richard & Miles & $2$ \\
%\bottomrule
%\end{tabular}
%\end{table}

%------------------------------------------------

\section{Closing}

In closing, we are satisfied with the degree of accomplishment of the objectives we set out from the start, namely disseminating knowledge on complex networks and emphasizing the importance of mathematics both for science and for our daily lives. We will most likely repeat the activity in the future working from the lessons we learned from these first attempts, particularly in the small group formats. We believe that there is material for divulgative lectures in larger format but it has be reworked to condensate it and make it more to the point in a shorter time. In this respect, it is worth mentioning two tools that came out during the preparation of the activities (actually, after we ran the small group sessions):
\begin{compactitem}
\item A website where strategies for vaccination and quarantining in social networks can be tried: {\tt http://vax.herokuapp.com/} This web can do precisely what we intented to do in this part (and that we never were able to get to). 
\item A short video on the effect of small changes on complex ecosystems: {\tt http://tinyurl.com/onoktvl}. We tried this video to finish the lecture at the High School and it was an instant hit. Aside from making extremely clear the need for understanding ecological networks, it points beyond the topic of complex networks and allows to connect with complex systems. 
\end{compactitem}
Finally, as \cite{mason} did, we encourage people to use, borrow, adapt, and improve any material in this article including the presentations available on the web. We are happy to advice anybody wanting to use the material and we would very much like hearing from you if you ever use it. 

%----------------------------------------------------------------------------------------
%	REFERENCE LIST
%----------------------------------------------------------------------------------------

%----------------------------------------------------------------------------------------

\end{multicols}

\end{document}